\documentclass[superscriptaddress,aps,preprintnumbers,amsmath,showpacs,amssymb,prd,nofootinbib,preprint]{revtex4-1}
\pdfoutput=1
\usepackage{booktabs} 
\usepackage{array} 
\usepackage[table,xcdraw]{xcolor} 
\usepackage{amsmath,amssymb}
\usepackage{graphicx}
\usepackage{fancyhdr}
\usepackage{bm, color}
\usepackage{slashed,amsthm,amsfonts,empheq}
\usepackage[caption=false]{subfig}
\usepackage{hyperref}
\usepackage{booktabs}
\usepackage{multirow}
\usepackage{natbib}
\usepackage{ulem}
\newcommand{\Slash}[1]{{\ooalign{\hfil#1\hfil\crcr\raise.167ex\hbox{/}}}}

\newcommand{\beq}{\begin{equation}}  \newcommand{\eeq}{\end{equation}}
\newcommand{\bef}{\begin{figure}}  \newcommand{\eef}{\end{figure}}
\newcommand{\bec}{\begin{center}}  \newcommand{\eec}{\end{center}}

\newcommand{\Sec}[1]{Sec.\ref{chap:#1}}

\newcommand{\lac}[1]{\label{chap:#1}}

\def\({\left(}
\def\){\right)}

\def\O{\mathcal{O}}

\def\f{\phi}

\def\*{\dagger}

\begin{document}
%
%
%
%
\preprint{}

\title{Diversity of Galaxy Centers from Small-Scale Isocurvature}


\author{Jessica N. Lopez-Sanchez}
\affiliation{CEICO—FZU, Institute of Physics of the Czech Academy of Sciences, Na Slovance 1999/2, 182 00 Prague, Czech Republic}

 \author{Wen Yin}
\affiliation{Department of Physics, Tokyo Metropolitan University, Minami-Osawa, Hachioji-shi, Tokyo 192-0397, Japan}

\begin{abstract}

The observed diversity of galaxy centers motivates the possibility that the local dark matter composition may vary among galaxies. In conventional multicomponent dark matter scenarios, however, large stochastic variations in the relative abundances are not generically expected, with the cold component typically remaining dominant. We perform $N$-body simulations with a subdominant ultralight dark matter component carrying significant large-amplitude small-scale isocurvature perturbations. We find that although nonlinear evolution largely homogenizes its fraction on galactic scales,ultralight dark matter can still be enhanced and even dominate the centers of small halos because large initial ultralight dark matter fluctuations seed early potential wells that later form halo centers. The ultralight-dominated region can extend beyond $0.01R_{\rm vir}$ and account for more than half of the central dark matter density, even for a cosmological ultralight dark matter fraction below $O(10\%)$. This central segregation provides a possible route to stochastic cusp--core diversity, potentially explaining the observed diversity of galaxy centers.

\noindent
\end{abstract}
\maketitle

\section{Introduction}

The cold dark matter (CDM) paradigm successfully accounts for the formation and distribution of cosmic structures over a broad range of scales. On galactic and subgalactic scales, however, observations reveal considerable diversity in the inferred central dark matter (DM) distributions. Galaxies with similar halo masses or maximum rotation velocities can exhibit markedly different inner rotation curves\cite{Oman:2015xda}. Whether the observed diversity is entirely due to baryonic processes and dynamical modeling, or instead reflects nontrivial properties of DM, therefore remains an open question~\cite{1601.05821,1611.02716,Bullock:2017xww,1808.05695,2202.00012,Nadler:2023nrd}. 
Dwarf spheroidal galaxies provide a complementary probe, with dynamical analyses indicating a diversity of inferred central DM profiles, while the profiles of ultra-faint dwarfs remain much more uncertain due to their limited stellar kinematic data \cite{2007.13780,2206.02821}.

A possible explanation is that DM consists of more than one component. In particular, a warm or ultralight dark matter (ULDM) component can suppress structure below a characteristic scale and, in the ultralight case, produce cored density profiles through wave effects \cite{Hu:2000ke,Hui:2016ltb,Marsh:2015xka}. However, scenarios in which warm or ultralight DM dominates the cosmological DM abundance are strongly constrained over much of the relevant parameter space by cosmological and astrophysical observations, including the Lyman-$\alpha$ forest, the abundance of small halos, and the formation of high-redshift galaxies \cite{Irsic:2017yje,Rogers:2020ltq,Winch:2024mrt}. A mixed scenario may evade these constraints. For such a scenario to generate substantial diversity through variations in the DM composition, however, the local component fractions must differ among galaxies, so that a cosmologically subdominant component dominates some galaxies, or some regions within them, while CDM remains dominant on average.

In conventional multicomponent DM scenarios with adiabatic initial conditions, all components trace the same primordial curvature perturbation. Although their later distributions can differ because of free streaming, wave effects, or self-interactions, large stochastic halo-to-halo variations in the component fractions, especially a subdominant component dominating only some halo centers, are not generically expected.

The key observation of this work is that component-specific isocurvature perturbations can provide a spatial separation of dark matter components, which could be a natural scenario. A light scalar field typically acquires isocurvature perturbations from inflation \cite{Seckel:1985tj,Lyth:1989pb,Lyth:1991ub} which is tightly constrained from cosmic-microwave background (CMB) data ~\cite{Planck:2018jri}. Small-scale isocurvature perturbations, which are much less constrained~\cite{Lee:2021bmn,Buckley:2025zgh}, may also arise. For example, they can originate from inflationary fluctuations of a field whose effective mass changes during inflation \cite{Kitano:2023mra}. Small fluctuations can also be enhanced through nonlinear evolution in the early Universe relevant to dark matter production~\cite{Gorghetto:2020qws, Kitajima:2023kzu,Yin:2024pri,Gorghetto:2025uls, Miyazaki:2025tvq,Masubuchi:2026eau,Aburatani:2026rct, Kitajima:2026llw}. The subdominant dark matter itself may consist of macroscopic objects with a small number density~\cite{astro-ph/0302035,1806.10414,2010.06470,2105.08462,2404.13110}. Alternatively it may be a very light field with a very long correlation length that is produced stochastically \cite{Amin:2022nlh,2510.15046,2510.17977}. Such stochastic behavior itself acts as an isocurvature perturbation.

In this paper, we investigate this possibility using cosmological $N$-body simulations containing a dominant CDM component and a subdominant ULDM component with isocurvature perturbations. We focus on the spatial segregation of the two components generated during nonlinear structure formation. We find that nonlinear evolution largely homogenizes the ULDM fraction when averaged over an entire halo. Nevertheless, halos seeded by sufficiently large ULDM isocurvature fluctuations can retain central regions dominated by ULDM, even though the halo as a whole remains CDM dominated. For the parameters studied here, the ULDM-dominated region can extend beyond $0.01R_{\rm vir}$ and contribute more than half of the central DM density, while the cosmological ULDM fraction remains below $\O(10\%)$. 

Mixed CDM--ULDM cosmologies have previously been studied in nonlinear simulations with adiabatic initial conditions, with particular emphasis on wave effects, halo abundance, and the dependence of core formation on the cosmological ULDM fraction \cite{Marsh:2013ywa,Marsh:2015wka,2007.08256,Lague:2023wes,2409.11469}.
Cosmological wave simulations of mixed cold and ultralight dark matter
with adiabatic primordial perturbations have found that, for a cosmological
ultralight fraction of $50\%$, the ultralight component dominates the inner
region of the halo and forms a solitonic core, whereas no stable core was
found for a fraction of $10\%$ \cite{Lague:2023wes}.
More recently, mixed warm and cold dark matter with a blue-tilted CDM isocurvature mode was investigated using $N$-body simulations, focusing on the nonlinear matter power spectrum and halo mass function \cite{2508.03805}. In contrast, the present work focuses on the radial segregation of the components within individual halos. In particular, we show that the global halo composition can be substantially homogenized while the central region retains a stochastic memory of the component-specific isocurvature seed. Some halos may develop central ULDM fractions exceeding $50\%$, which may produce central cores there following \cite{Lague:2023wes}, although  the cosmological ULDM fraction is below $\O(10\%)$.

 Our results provide a mechanism that could potentially solve the diversity problem: galaxies sharing a similar cosmological DM composition can acquire qualitatively different central compositions, i.e., the ULDM inner halo could form a cored profile.

The remainder of this paper is organized as follows. We first discuss early-Universe scenarios for the isocurvature perturbations and the relevant observational limits in \Sec{2}. The two-component initial conditions and describe the numerical simulation pipeline are introduced in \Sec{3}. After presenting the resulting halo and inner-halo diversity in \Sec{4}, we discuss the implications and summarize our conclusions.

\section{Early Universe Scenario for Isocurvatures and Limits}
\lac{2}
To motivate our numerical setup, we consider a dark sector containing two
dark matter (DM) components, although the formalism can be straightforwardly
extended to more than two components.

The first component is the dominant cold dark matter (CDM), while the second
is a cosmologically subdominant component. In our numerical examples, we
take the latter to be ultralight dark matter (ULDM) and denote quantities
associated with CDM and ULDM by the subscripts $c$ and $u$, respectively.
The component-segregation mechanism studied below is, however, not specific
to ULDM. At the level of the initial conditions, it only requires that the
subdominant component carry an independent small-scale fluctuation that
survives until halo formation. A warm, self-interacting, or otherwise
non-CDM component is also interesting.\footnote{For example, the subdominant component may belong to a
decoupled dark sector and have a thermal history different from that of the
Standard Model sector. Such a component can carry isocurvature perturbations
if its local abundance or temperature is controlled by a degree of freedom
or production process independent of the inflaton.}

We denote the primordial isocurvature field associated with the
subdominant component by $S(\mathbf{k})$ and define its dimensionless power
spectrum by
\beq
\left\langle
S(\mathbf{k})S^*(\mathbf{k}')
\right\rangle
=
(2\pi)^3
\delta_{\rm D}^{(3)}(\mathbf{k}-\mathbf{k}')
\frac{2\pi^2}{k^3}
\mathcal{P}_{SS}(k).
\eeq
The isocurcature can be not only be generated through the inflationary flucutation, 
but also from the stochastic process for a late time cosmology. 
In either case one can  have significantly large isocurvature if the cosmology undergo a nonlinear evolution, e.g. the ULDM is produced from the collapse topological defects or long enough corellation length.


We first considered a scale-invariant primordial isocurvature spectrum,
\begin{equation}
    \mathcal{P}_{SS}(k)
    =
    A_s \beta
    \left(\frac{k}{k_*}\right)^{n_s-1},
\end{equation}
where $\beta$ denotes the ratio
$\mathcal{P}_{SS}(k_*)/\mathcal{P}_{RR}(k_*)$.
Here, $A_s$ is the amplitude of the primordial adiabatic curvature
spectrum at the pivot scale $k_*$.

In this case, we find no significant impact of primordial isocurvature
perturbations satisfying the stringent limits from CMB
data~\cite{Planck:2018jri} on the final relative distribution of the two
dark matter components within halos. Instead, CDM and ULDM efficiently
populate the same gravitational potential wells, and the halo composition
remains close to the cosmic abundance of each component. This indicates
that gravitational evolution efficiently mixes the two components and
prevents the formation of a distinct population of compositionally
segregated halos. These conclusions will be revisited in the future
(see \Sec{Discussion}).

In this paper, therefore we consider the impact of the small scale and parametrize the power spectrum with two choices: localized 
\beq
\mathcal{P}_{SS}^{\rm NB}(k)
=
A_s\beta_{\rm NB}
\delta\!\left(
\ln k-\ln k_0
\right).
\label{eq:iso_narrow_band}
\eeq
and 
the smooth
broken-power-law template
\beq
\mathcal{P}_{SS}^{\rm BPL}(k)
=
A_s\beta_{\rm BPL}
\frac{1}
{\left(k/k_0\right)^{-3}+(k/k_0)^{p}}.
\label{eq:iso_broken_power_law}
\eeq
In the Appendix.~\ref{app:1}, we introduce scenarios for the different spectral templates. 

The observational effect of an isocurvature mode carried by a subdominant
matter component is suppressed by its fractional contribution to the
total matter density. We define
\beq
r
\equiv
\frac{\Omega_u}{\Omega_{\rm DM}},
\qquad
r_m
\equiv
\frac{\Omega_u}{\Omega_m}
=
r\frac{\Omega_{\rm DM}}{\Omega_m}.
\eeq
Here $\mathcal{P}_{SS}$ is defined as the intrinsic isocurvature spectrum
of the $u$ component. The contribution to cosmological observables
therefore scales approximately as
\beq
r_m^2\mathcal{P}_{SS}.
\eeq

The limits are substantially weaker when the isocurvature power is
localized on small scales. Ref.~\cite{Buckley:2025zgh} derives
constraints on both the narrow-band and broken-power-law templates using
CMB anisotropies, baryon acoustic oscillations, the Lyman-$\alpha$ forest,
and CMB spectral distortions. Identifying their isocurvature amplitude
with $A_{\rm iso}=A_s\beta$, the current bounds shown,
at the order-of-magnitude level,
\beq
\beta_{\rm BPL}
\lesssim
\mathcal{O}
\left(10^3\text{--}10^6\right)
r_m^{-2},
\qquad
10\,{\rm Mpc}^{-1}
\lesssim
k_0
\lesssim
100\,{\rm Mpc}^{-1},
\label{eq:bpl_iso_limit}
\eeq
and
\beq
\beta_{\rm NB}
\lesssim
\mathcal{O}\left(10^{11}\text{--}10^{12}\right)r_m^{-2},
\qquad
10\,{\rm Mpc}^{-1}
\lesssim
k_0
\lesssim
100\,{\rm Mpc}^{-1}.
\label{eq:narrow_iso_limit}
\eeq
Thus, $\beta_{\rm BPL}\lesssim\mathcal{O}(10^5)r_m^{-2}$ may be used as
a representative estimate within this interval, but it is not a
$k_0$-independent bound. The broken-power-law constraint at
$k_0\gg{\rm Mpc}^{-1}$ is obtained by extrapolating the
Lyman-$\alpha$ result with its approximate $k_0^3$ scaling. For the
narrow-band spectrum, the current constraints in this range arise mainly
from the COBE/FIRAS limits on $y$- and $\mu$-type spectral distortions.
The projected PIXIE sensitivity is 
$
\beta\sim
\mathcal{O}\left(10^8\text{--}10^9\right)r_m^{-2}
$
over part of the same range. Thus the focus of our study may be probed in the future. 

In the following we introduce the isocurvature satisfying the limits given above. 

\section{Setup for N-body simulation}\lac{3}
Motivated by the previous disucussion, we perform a two component N-body simulation by taking one compoenent with the isocurvaure fluctuation. 
\subsection{Initial conditions}
\label{sec:initial_conditions}
We consider a two-component dark matter model composed of CDM, denoted by $c$, and ULDM, denoted by $u$. In the linear regime, the Fourier-space density contrasts of the two components can be written as
\begin{align}
\delta_c(\mathbf{k})
&=
T_c^{R}(k)\,R(\mathbf{k})
+
T_c^{S}(k)\,S(\mathbf{k}),
\\
\delta_u(\mathbf{k})
&=
T_u^{R}(k)\,R(\mathbf{k})
+
T_u^{S}(k)\,S(\mathbf{k}),
\end{align}
where
$
\displaystyle\delta_i(\mathbf{x})
=
\frac{\rho_i(\mathbf{x})-\bar{\rho}_i}
{\bar{\rho}_i}
$ is the density contrast of component $i$.

The primordial fields $R(\mathbf{k})$ and $S(\mathbf{k})$ denote the adiabatic and isocurvature perturbations, respectively, while $T_i^R(k)$ and $T_i^S(k)$ are the corresponding transfer functions, computed with the public Boltzmann code \texttt{axionCAMB}~\cite{Hlozek:2014lca,hlovzek2018using}. {We adopt $m_u=10^{-22}\,{\rm eV}$ as a fiducial ULDM mass when computing the linear transfer functions\footnote{%
Bounds on $m_u$ for a given $f_u$, derived for adiabatic initial
conditions~\cite{Bozek:2014uqa,Kobayashi:2017jcf,Bar:2021kti,Liu:2026jkq},
do not directly apply to large small-scale isocurvature, which instead
enhances small-scale structure (see Fig.~\ref{fig: projected_densities}).
Our goal is to demonstrate matter segregation rather than construct a
fully viable cosmology. A similar mechanism may operate in single-component
dark matter if baryonic isocurvature generates spatial variations in the
baryon-to-dark-matter ratio and hence in galaxy central composition.%
}

In the subsequent $N$-body evolution, both dark matter components are treated as collisionless particles. Thus, in this work $m_u$ sets the linear small-scale suppression of the ULDM transfer function, while our main goal is to isolate how localized primordial isocurvature modifies the relative CDM/ULDM composition of halos, which is only weakly affected by this suppression.

Throughout this work, we assume that the primordial adiabatic and isocurvature perturbations are statistically independent,
$\left\langle R(\mathbf{k})S^*(\mathbf{k}') \right\rangle = 0$,
or equivalently $P_{RS}(k)=0$.
Only the ULDM component carries a primordial isocurvature perturbation, while the CDM component is purely adiabatic, although the two components subsequently mix through their gravitational interaction.

For the simulations presented below, the adiabatic spectrum is fixed to
  \begin{equation}
  \mathcal{P}_{RR}(k)
  =
  A_s
  \left(\frac{k}{k_*}\right)^{n_s-1},
  \end{equation}
  with $A_s=2.196\times10^{-9}$, $n_s=0.9655$, and
  $k_*=0.05\,{\rm Mpc}^{-1}$. 
  The corresponding dimensional primordial power spectra are
$P_R(k)
=
\displaystyle\frac{2\pi^2}{k^3}\,
\mathcal{P}_{RR}(k)$ and for the isocurvature 
$P_S(k)
=
\displaystyle\frac{2\pi^2}{k^3}\,
\mathcal{P}_{SS}(k).
$


\subsection{Numerical pipeline}
\label{sec:numerical_realization}
Assuming that the primordial adiabatic and isocurvature perturbations
are uncorrelated, $P_{RS}(k)=0$, the linear auto- and cross-power
spectra of the two dark-matter components are
\begin{align}
P_{cc}(k)
&=
|T_c^R(k)|^2P_{RR}(k)
+
|T_c^S(k)|^2P_{SS}(k),
\\
P_{uu}(k)
&=
|T_u^R(k)|^2P_{RR}(k)
+
|T_u^S(k)|^2P_{SS}(k),
\\
P_{cu}(k)
&=
T_c^R(k)T_u^{R*}(k)P_{RR}(k)
+
T_c^S(k)T_u^{S*}(k)P_{SS}(k).
\end{align}
These spectra include the response of both dark-matter components to
the primordial adiabatic and isocurvature modes.

The physical model is implemented numerically through the covariance
matrix of the CDM and ULDM density fields,
\begin{equation}
\mathbf{P}(k)
=
\begin{pmatrix}
P_{cc}(k) & P_{cu}(k)\\
P_{cu}^*(k) & P_{uu}(k)
\end{pmatrix},
\end{equation}
which is tabulated and supplied as input to our modified version of
\texttt{N-GenIC}\cite{Springel:2005nw}. For each Fourier mode, the covariance matrix is
factorized through a Cholesky decomposition,
\begin{equation}
\mathbf{P}(k)
=
\mathbf{L}(k)\mathbf{L}^{\dagger}(k),
\qquad
\mathbf{L}(k)
=
\begin{pmatrix}
L_{11}(k) & 0\\
L_{21}(k) & L_{22}(k)
\end{pmatrix},
\end{equation}
with
\begin{align}
L_{11}(k)
&=
\sqrt{P_{cc}(k)},
\\
L_{21}(k)
&=
\frac{P_{cu}^*(k)}{L_{11}(k)},
\\
L_{22}(k)
&=
\left[
P_{uu}(k)-|L_{21}(k)|^2
\right]^{1/2}.
\end{align}
The vanishing upper-right entry is not an additional physical
assumption, but follows from the choice of a lower-triangular Cholesky
factor. In particular, the response of CDM to the primordial
isocurvature mode is fully included through $T_c^S$ in $P_{cc}$ and
$P_{cu}$, while the CDM--ULDM cross-correlation is encoded in
$L_{21}$.

Two independent, unit-variance complex Gaussian random fields,
$g_1(\mathbf{k})$ and $g_2(\mathbf{k})$, are then generated, and the
Fourier-space density fields are constructed as
\begin{align}
\delta_c(\mathbf{k})
&=
L_{11}(k)\,g_1(\mathbf{k}),
\\
\delta_u(\mathbf{k})
&=
L_{21}(k)\,g_1(\mathbf{k})
+
L_{22}(k)\,g_2(\mathbf{k}).
\end{align}
By construction, these fields satisfy
\begin{align}
\left\langle
\delta_c(\mathbf{k})\delta_c^*(\mathbf{k})
\right\rangle
&=P_{cc}(k),
\\
\left\langle
\delta_u(\mathbf{k})\delta_u^*(\mathbf{k})
\right\rangle
&=P_{uu}(k),
\\
\left\langle
\delta_c(\mathbf{k})\delta_u^*(\mathbf{k})
\right\rangle
&=P_{cu}(k).
\end{align}
The construction is well defined provided that the covariance matrix
is positive semidefinite,
\begin{equation}
P_{cc}(k)P_{uu}(k)-|P_{cu}(k)|^2\geq0.
\end{equation}
In the present implementation, the tabulated transfer functions and
power spectra are real, such that $P_{cu}^*=P_{cu}$ and
$\mathbf{L}^{\dagger}=\mathbf{L}^{\rm T}$. Particle velocities are initialized using the standard Zel'dovich
prescription implemented in \texttt{N-GenIC}.\footnote{Using \texttt{axionCAMB} transfer functions at nearby redshifts, we find
that scale-dependent growth differs from the standard \texttt{N-GenIC}
prefactor by at most ${\cal O}(10)\%$ on scales relevant for the initial
displacements, including $k\simeq k_0$. Modes with larger deviations are
negligible. Low-resolution tests also confirm that the initial-velocity
treatment does not significantly affect our results.}

The resulting particle distribution constitutes the initial conditions for the subsequent multi-component N-body simulations performed with multi-component \texttt{Gadget-4} \cite{springel2021simulating}. We adopt a flat $\Lambda$CDM cosmology with $\Omega_b=0.05$, $\Omega_\Lambda=0.70$, $h=0.67$, and initial redshift $z_{\rm ini}=127$. The CDM and ULDM density parameters, $\Omega_{\rm c}$ and $\Omega_{\rm u}$, are chosen such that $\Omega_{\rm c}+\Omega_{\rm u}=\Omega_{\rm DM}=0.25$. We consider a set of simulations spanning different values of $\beta_{\rm NB}$ and $\beta_{\rm BPL}$ for the narrow-band isocurvature spectrum as well as different cosmic fractions$f_u\equiv\Omega_u/\Omega_{\rm DM}$ for the broken-power law one. We consider $k_0=10\,{\rm Mpc}^{-1}$. Haloes are subsequently identified using our modified version of \texttt{Rockstar} \cite{behroozi2013rockstar}, which distinguishes between the CDM and ULDM components when constructing halo catalogues.

Figure~\ref{fig: projected_densities} illustrates the projected matter
distribution for different cosmic ULDM fractions in the Broken Power-Law
case only, as a representative example. While the large-scale cosmic web remains qualitatively similar,
increasing $f_u$ produces a more granular small-scale distribution, with
more prominent compact density enhancements along the filamentary
structure. This reflects the increasing contribution of ULDM
isocurvature perturbations to the total matter fluctuations. Thus, a
larger ULDM fraction does not simply enhance the usual small-scale
suppression, as the primordial isocurvature fluctuations can instead
modify the amount and distribution of small-scale structure. This behaviour will be discussed in detal in the following section.

Figure~\ref{fig: halo_snapshot} displays a representative halo with an ULDM-enhanced central region in the broken power-law simulation with $f_u=0.3$, as an ilustrative example. The large-scale environment shows that the two dark-matter components trace the same underlying gravitational structures. The zoomed-in view (middel and right panels) highlights the spatially localized halo, whereas the projected ULDM fraction demostrated that this enhancement is concentrated in the inner halo rather than corresponding to a globally ULDM-dominated object. This provides a direct visualization of the central ULDM enrichment quantified statistically in Figs.~\ref{fig: delta_rvir} and~\ref{fig: brokenlaw_concentration}, later discussed.

\begin{figure}
    \centering
    \includegraphics[width=\linewidth]{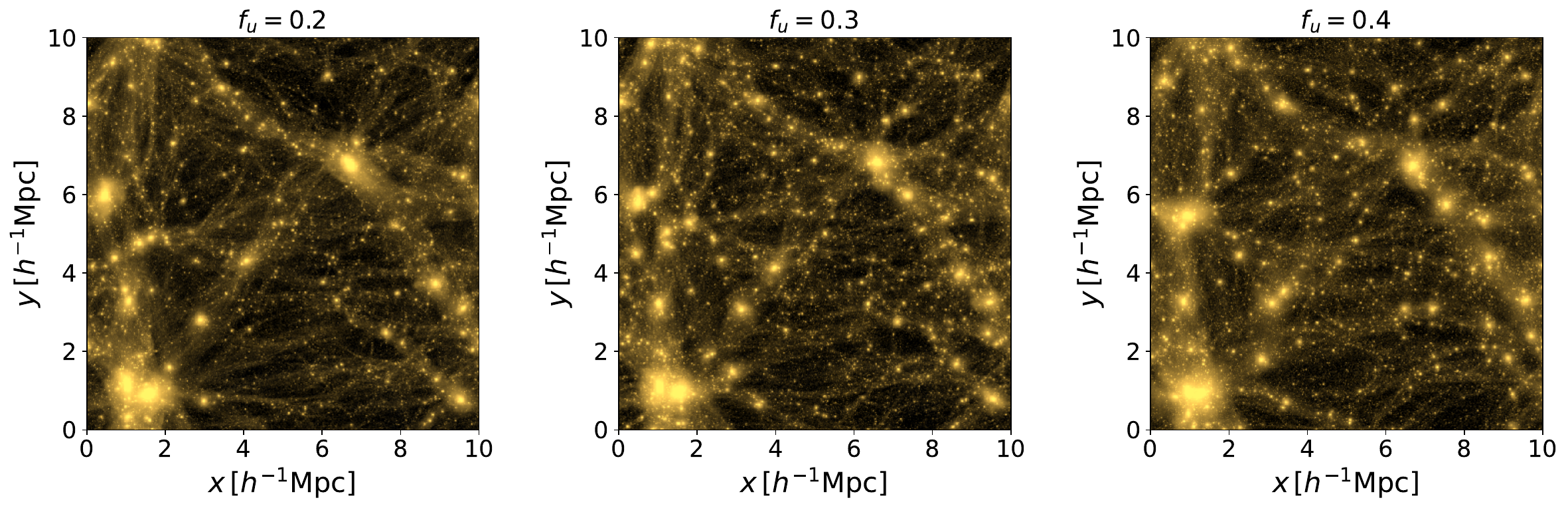}
    \caption{Projected matter density for the Broken Power-Law isocurvature
simulations with different cosmic ULDM fractions, $f_u=0.2$, $f_u=0.3$ and $0.4$. The projections show the same region of the simulation volume, illustrating the preservation of the large-scale filamentary structure and the changes in the small-scale matter distribution as the ULDM fraction is increased. Brighter regions correspond to larger projected matter densities.}
    \label{fig: projected_densities}
\end{figure}

\begin{figure}
    \centering
 \includegraphics[width=\linewidth]{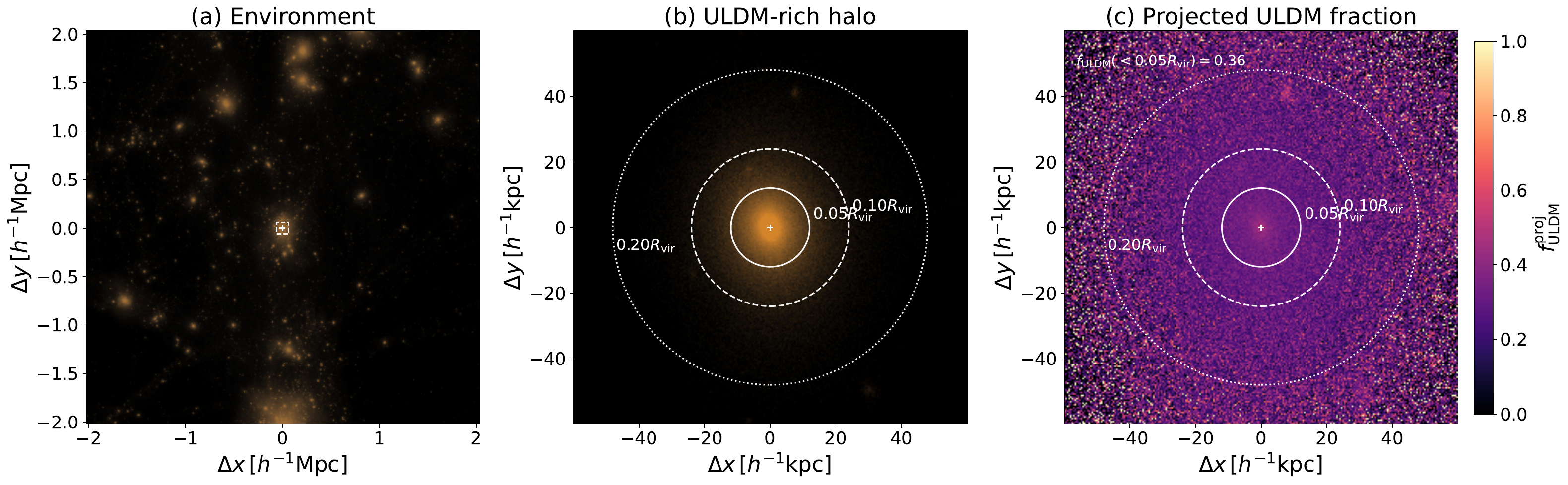}
    \caption{Example of a halo with an ULDM-enhanced central region in the broken power-law simulation with $f_u=0.3$. Left: projected matter distribution in the large-scale environment of a selected halo. Middle: zoomed-in of the selected halo. Right: projected ULDM fraction in the same region. The circles indicate $0.05R_{\rm vir}$, $0.1R_{\rm vir}$, and $0.2R_{\rm vir}$. }
    \label{fig: halo_snapshot}
\end{figure}

\section{Halo and Inner Halo Diversity from isocurvature}
\lac{4}
To characterize the relative contribution of ULDM to each dark matter halo, we define the halo ULDM mass fraction as

\begin{equation}
f_{\rm ULDM}
=
\frac{M_u}{M_c+M_u},
\end{equation}
where $M_c$ and $M_u$ denote the total CDM and ULDM masses assigned to the halo, respectively. In practice, these masses are obtained by summing the particle masses of each species belonging to the same \texttt{Rockstar} halo. Throughout this work, halos with $f_{\rm ULDM}>0.5$ are classified as ULDM-dominated, whereas values above the cosmic ULDM fraction indicate a local enhancement of ULDM
relative to the cosmological mean.

The radial distribution of ULDM within halos is measured using the following expression
\begin{equation}
f_{\rm ULDM}(<xR_{\rm vir})
=
\frac{M_u(<xR_{\rm vir})}
{M_c(<xR_{\rm vir})+M_u(<xR_{\rm vir})},
\end{equation}
where $x=0.01$, $0.05$, and $0.1$.

Finally, to quantify the degree of central ULDM enhancement independently of the overall halo composition, we define the central enhancement factor,
\begin{equation}
\mathcal{C}_x
\equiv
\frac{f_{\rm ULDM}(<xR_{\rm vir})}
     {f_{\rm ULDM}},
\end{equation}
where $\mathcal{C}_x>1$ indicates that the inner region is more ULDM-rich than the halo as a whole, while $\mathcal{C}_x=1$ corresponds to the same ULDM fraction in the inner region and the halo.

\subsection{Narrow-Band spectrum}

Figure~\ref{fig:beta_tot_histogram} shows the evolution of the halo ULDM mass-fraction distribution for the narrow-band isocurvature runs.\footnote{We choose $\beta$ to satisfy the limit in Eq.~\eqref{eq:narrow_iso_limit}. For sufficiently large $\beta$, however, the scalar-field perturbations become nonlinear already at very high redshift, invalidating the usual linear treatment and potentially modifying the small-scale halo distribution. A dedicated study of this regime and its implications for isocurvature constraints is left for future work.} At early times, the distributions span a broad range of $f_{\rm ULDM}$, indicating that localized primordial isocurvature fluctuations seed halos with diverse initial ULDM compositions. As structure formation proceeds, the bulk of the distributions progressively shifts toward smaller $f_{\rm ULDM}$, reflecting the increasing contribution of CDM to the same gravitational potential wells. Consequently, the abundance of ULDM-dominated halos decreases with time, showing that the initial ULDM enhancement is gradually diluted during nonlinear evolution. Even when we adopt very large initial ULDM fluctuations, nonlinear evolution strongly suppresses ULDM-dominated halos, with halos satisfying $f_{\rm ULDM}>0.5$ constituting only a small fraction of the halo population around the present epoch. Nevertheless, halo-to-halo variations persist, indicating that primordial isocurvature leaves a lasting imprint on the diversity of halo compositions.

\begin{figure}
    \centering
    \includegraphics[width=1\linewidth]{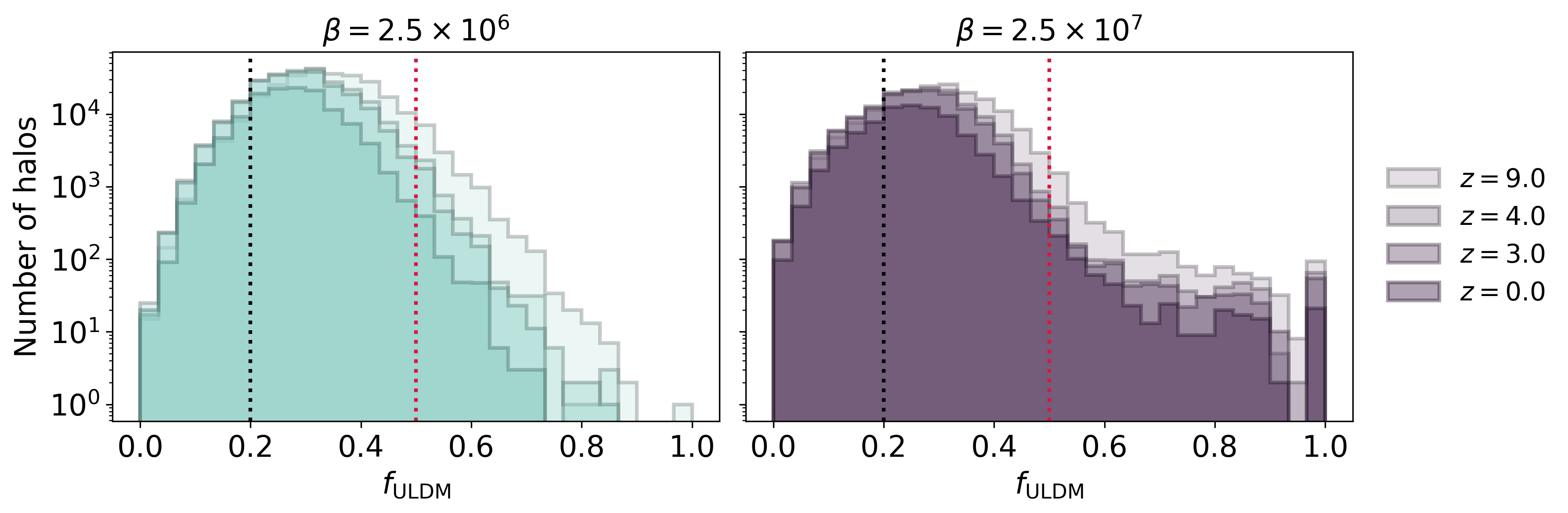}
    \caption{
     Evolution of the halo ULDM mass-fraction distribution for the narrow-band isocurvature simulations with different isocurvature amplitudes with $f_u=0.2$. Histograms show the distribution of $f_{\rm ULDM}$ at different redshifts for the $\beta=2.5\times10^6$ and $\beta=2.5\times10^7$ models. The black dashed line marks the corresponding cosmic ULDM fraction, while the red dotted line indicates $f_{\rm ULDM}=0.5$. In both models, the distributions shift toward smaller values of $f_{\rm ULDM}$ with time, reducing the abundance of ULDM-dominated halos.
}
    \label{fig:beta_tot_histogram}
\end{figure}

To explore whether this halo-to-halo diversity extends to the inner regions of halos, we measure the ULDM mass fraction within fixed fractions of the virial radius. Figure~\ref{fig: beta_rvir} compares the total halo ULDM fraction with $f_{\rm ULDM}(<xR_{\rm vir})$ for $x=0.01$, $0.05$, and $0.1$. Halos above the diagonal contain a larger ULDM fraction in their central regions than in the halo as a whole, whereas halos below the diagonal exhibit a more extended ULDM distribution. This central enhancement may arise because regions seeded by large initial ULDM fluctuations collapse first and become incorporated into the central parts of halos, while the surrounding CDM subsequently falls into the same gravitational potential wells.
{The broad scatter demonstrates that halos with similar global ULDM fractions can develop different central compositions.}

\begin{figure}
    \centering
    \includegraphics[width=\linewidth]{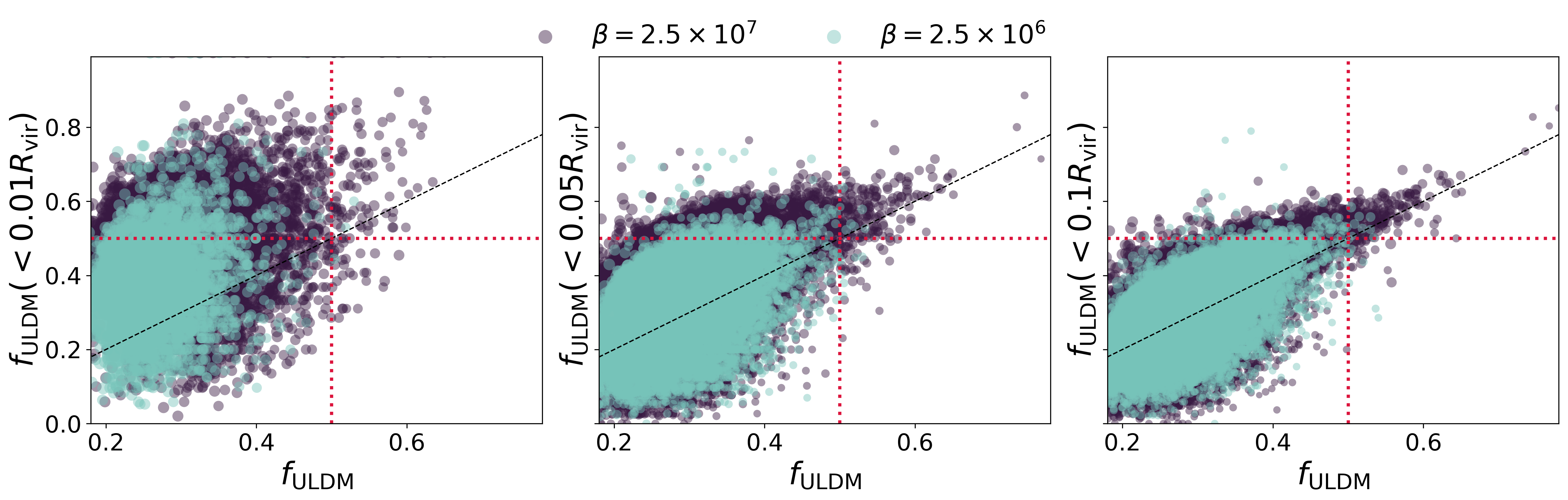}
    \caption{ULDM mass fraction measured within fixed fractions of the virial radius as a function of the total halo ULDM fraction $f_{\rm ULDM}$ for narrow band spectra with $f_u=0.2$,  at $z=0$. Each panel shows a different aperture, $0.01R_{\rm vir}$, $0.05R_{\rm vir}$, and $0.1R_{\rm vir}$ for $\beta=2.5\times10^6$ and $\beta=2.5\times10^7$. The horizontal and vertical red dotted lines mark $f_{\rm ULDM}=0.5$. Points above the diagonal are more ULDM-enriched in their inner regions. Marker size scales with the halo virial mass, $M_{\rm vir}$, using a common normalization across all panels. We include halos with at least 300 particles in total and at least 30 particles within the corresponding aperture. }
    \label{fig: beta_rvir}
\end{figure}

Figure~\ref{fig: beta_rvir_Mvir} shows the ULDM fraction measured within different fractions of the virial radius as a function of halo virial mass. The spread in the central ULDM fraction is substantially larger for low-mass halos than for high-mass halos, with the distribution becoming progressively narrower toward increasing halo mass. While the most massive halos exhibit relatively similar central ULDM fractions, low-mass halos span a much broader range of values. {This trend is consistent with the hierarchical assembly of dark matter halos, in which repeated mergers, continuous accretion, and the associated phase-space mixing progressively erase the memory of the initial ULDM enrichment, leading to more homogeneous central ULDM fractions in massive halos \cite{wang2011assembly,fakhouri2010merger,helmi2003phase},} {and the enhanced fluctuation only around the small scale.}

\begin{figure}
    \centering
    \includegraphics[width=\linewidth]{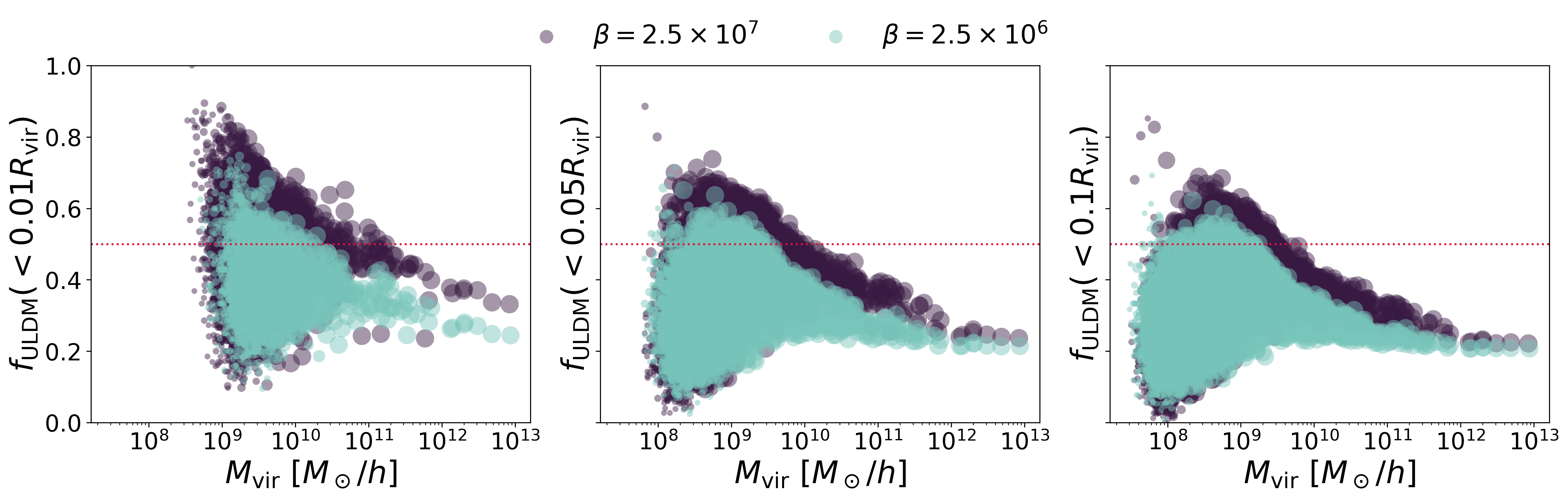}
    \caption{Inner ULDM mass fraction as a function of halo virial mass using the narrow band spectra with $f_u=0.2$ at $z=0$. Each panel shows the ULDM fraction measured within the same apertures as in Fig. \ref{fig: beta_rvir}. The red dotted line marks $f_{\rm ULDM}=0.5$ as before. Marker size indicates the number of particles within the corresponding aperture.}
    \label{fig: beta_rvir_Mvir}
\end{figure}

Figure~\ref{fig:  beta_concentration} summarizes these results through the survival distribution of the central enhancement factor, $\mathcal{C}_x$, providing a statistical characterization of the diversity of central ULDM enrichment across the halo population. While the median central enhancement changes only moderately, from $\mathcal{C}_{0.01}=1.44$ to $1.55$, the abundance of strongly enhanced systems changes much more significantly. Within $0.05R_{\rm vir}$, the fraction of halos with $\mathcal{C}_x>2$ increases from $0.16\%$ to $5.0\%$, corresponding to more than an order-of-magnitude increase. These results show that increasing the primordial isocurvature amplitude primarily broadens the distribution of central ULDM enhancement, boosting the incidence of ULDM-rich galaxy centers rather than uniformly increasing the central ULDM fraction across the halo population.

\begin{figure}
    \centering
    \includegraphics[width=\linewidth]{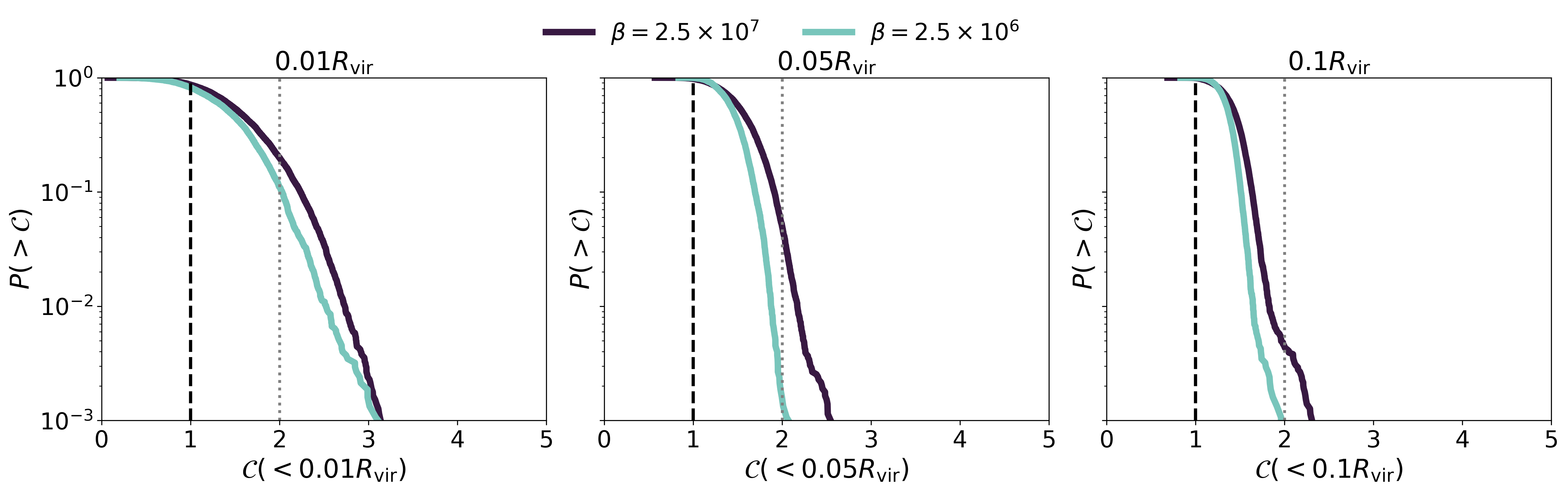}
    \caption{Survival distribution of the central enhancement factor, $\mathcal{C}_x$, measured within $0.01R_{\rm vir}$, $0.05R_{\rm vir}$, and $0.1R_{\rm vir}$ for the narrow-band isocurvature simulations with $f_u=0.2$ at $z=0$. The black dashed line marks $\mathcal{C}_x=1$, where the inner region has the same ULDM fraction as the halo as a whole, while the gray dotted line indicates $\mathcal{C}_x=2$. Larger isocurvature amplitudes produce a higher fraction of halos with large central enhancements.}
    \label{fig: beta_concentration}
\end{figure}

\subsection{Broken Power-Law Spectrum}

Since the broken-power-law spectrum is more strongly constrained, here we examine whether similar conclusions can be obtained with a slightly smaller $\beta$ and a larger $f_u$. We take $\beta=10^6$ as our fiducial value, although smaller values or larger $k_0$ also work provided that $f_u$ is not too small.

Figure~\ref{fig:delta_tot_histogram} shows the same analysis as Fig.~\ref{fig:beta_tot_histogram} for the broken power-law spectrum simulations with different cosmic ULDM fractions, while keeping the primordial amplitud $\beta$ fixed. As expected, increasing the cosmic ULDM fraction shifts the entire distribution toward larger values of $f_{\rm ULDM}$, producing a larger population of ULDM-rich halos at all redshifts. Despite this shift, all three models exhibit the same qualitative evolution: the distributions progressively move toward smaller halo ULDM fractions with time, and the abundance of ULDM-dominated halos decreases substantially by $z=0$. Comparing Figs.~\ref{fig:beta_tot_histogram} and \ref{fig:delta_tot_histogram} shows that the two parameters affect the halo population in different ways. Increasing the cosmic ULDM fraction primarily shifts the bulk of the distribution toward larger $f_{\rm ULDM}$, whereas increasing the primordial isocurvature amplitude mainly broadens the high-$f_{\rm ULDM}$ tail, increasing the abundance of the most ULDM-rich halos.

\begin{figure}
    \centering
    \includegraphics[width=\linewidth]{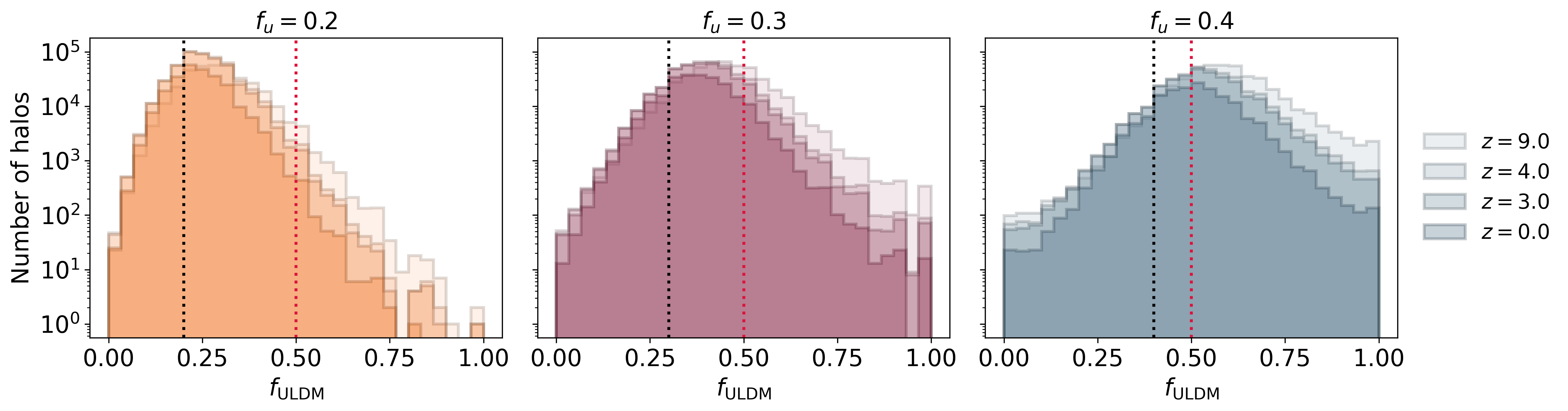}
    \caption{Evolution of the halo ULDM mass-fraction distribution for simulations with different cosmic ULDM fractions using with broken power-law spectrum with $\beta=10^6$. Histograms show the distribution of $f_{\rm ULDM}$ at different redshifts for the $f_{20}$, $f_{30}$, and $f_{40}$ models. The black dashed line marks the corresponding cosmic ULDM fraction of each simulation, while the red dotted line indicates $f_{\rm ULDM}=0.5$. In all cases, the distributions shift toward smaller values of $f_{\rm ULDM}$ with time, reducing the abundance of ULDM-dominated halos as structure formation proceeds. 
}
    \label{fig:delta_tot_histogram}
\end{figure}

Figure~\ref{fig: delta_rvir} presents the inner concentration of ULDM in these simulations. The same overall behaviour is recovered in all three models: although many halos exhibit enhanced ULDM concentrations in their inner regions, halos with similar total ULDM fractions can still develop different central compositions. Increasing the cosmic ULDM fraction systematically shifts the halo population toward larger values of both the total and central ULDM fractions, while preserving the broad correlation and scatter observed in Fig.~\ref{fig: beta_rvir}.

\begin{figure}
    \centering
    \includegraphics[width=\linewidth]{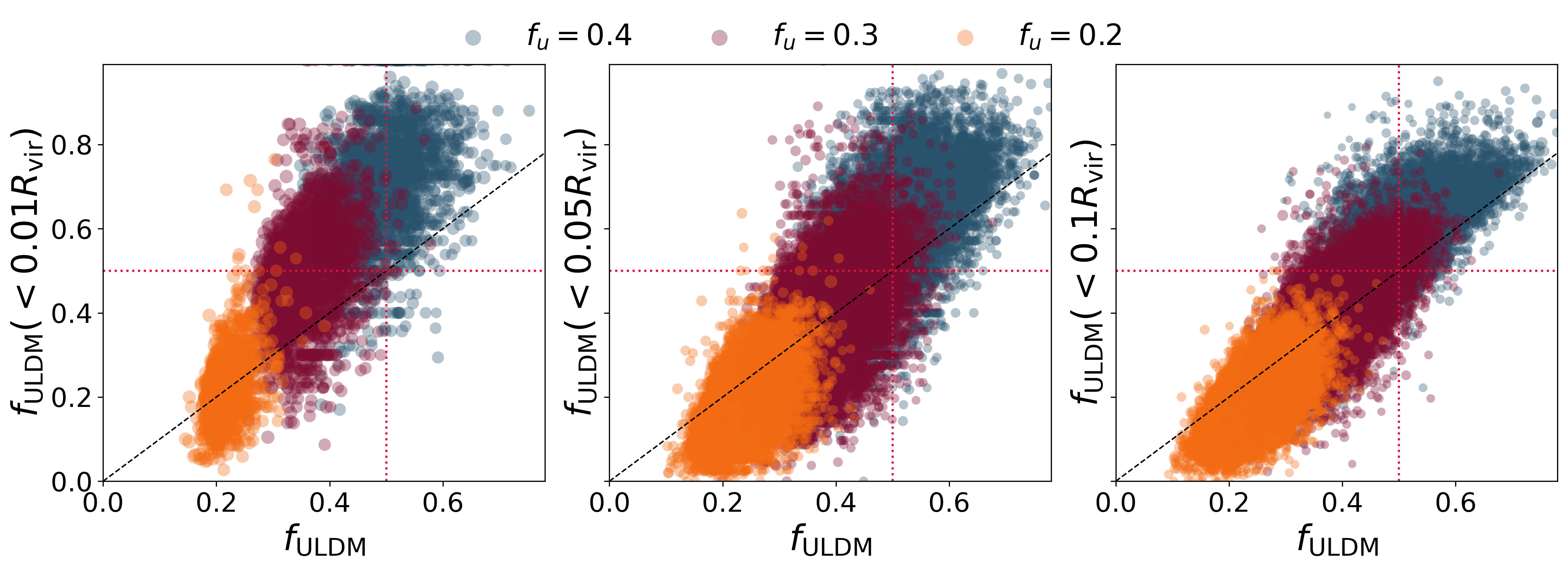}
    \caption{ULDM mass fraction measured within fixed fractions of the virial radius as a function of the total halo ULDM fraction $f_{\rm ULDM}$ at $z=0$ for simulations with different cosmic ULDM fractions. Here we consider broken power law spectrum with $\beta=10^6$. Each panel shows the same aperture for $R_{\rm vir}$ as in Fig.~\ref{fig: beta_rvir}. }
    \label{fig: delta_rvir}
\end{figure}

Figure~\ref{fig: delta_rvir_Mvir} shows the ULDM fraction measured within different fractions of the virial radius as a function of halo virial mass for simulations with different cosmic ULDM fractions. The same dependence on halo mass is recovered in all three models as in Fig.~\ref{fig: beta_rvir_Mvir}: low-mass halos exhibit a broad diversity of central ULDM fractions, whereas the distributions become progressively more homogeneous toward higher halo masses.

\begin{figure}
    \centering
    \includegraphics[width=\linewidth]{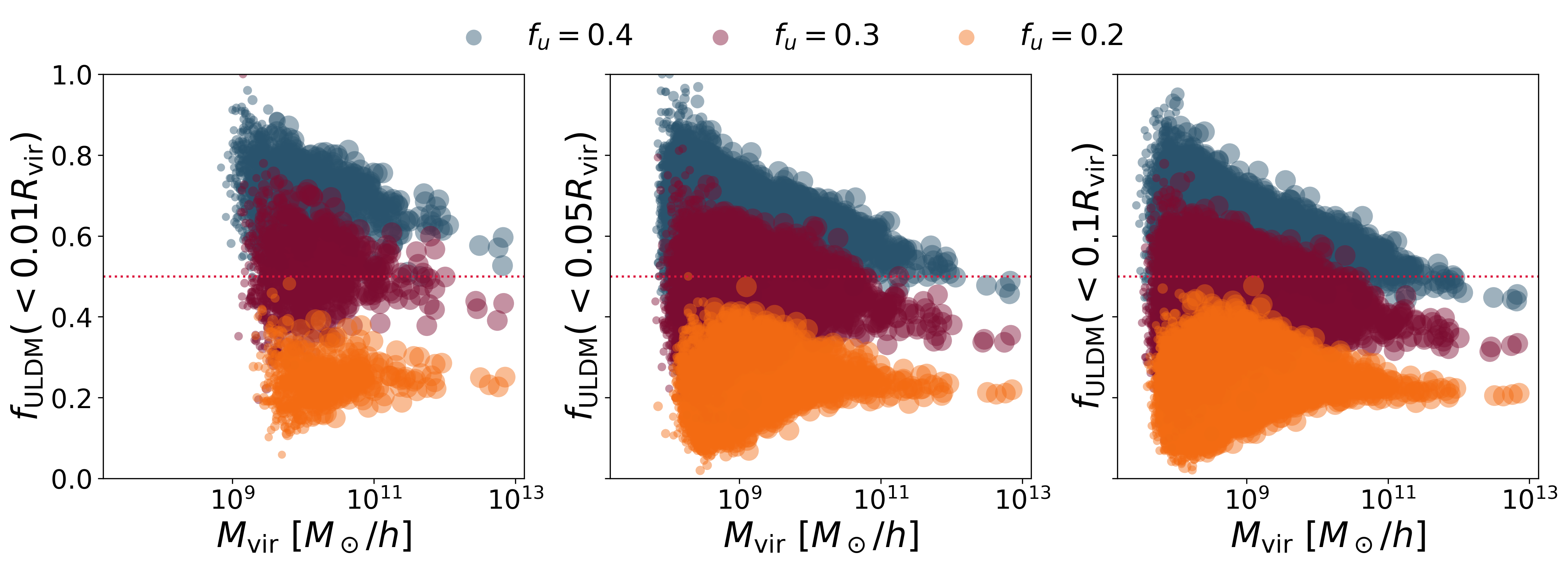}
    \caption{Inner ULDM mass fraction as a function of halo virial mass. Each panel shows the ULDM fraction measured within the same apertures as in Fig. \ref{fig: beta_rvir}. }
    \label{fig: delta_rvir_Mvir}
\end{figure}

Finally, Fig.~\ref{fig: brokenlaw_concentration} shows that, although increasing $f_u$ produces a larger population of globally ULDM-rich and even nearly pure-ULDM halos, it does not lead to a systematic increase in the relative central enhancement $\mathcal{C}_x$, indicating that the enhanced ULDM abundance affects primarily the overall halo composition rather than producing an increasingly stronger central segregation.

\begin{figure}
    \centering
    \includegraphics[width=\linewidth]{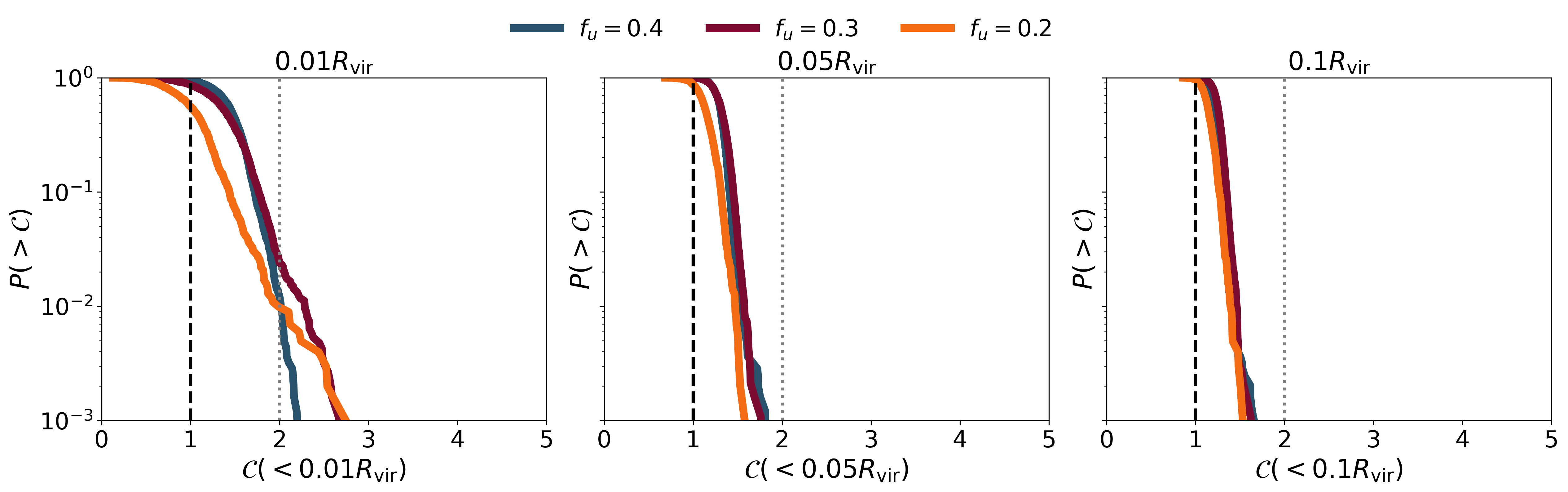}
    \caption{Survival distribution of $\mathcal{C}_x$, measured within $0.01R_{\rm vir}$, $0.05R_{\rm vir}$, and $0.1R_{\rm vir}$ for the broken power-law simulations with different cosmic ULDM fractions at $z=0$, with $\beta=10^6$. The dependence on the cosmic ULDM fraction is strongest at the smallest aperture, where larger $f_u$ generally increases the abundance of halos with enhanced central ULDM fractions, while the distributions become increasingly similar at larger radii. The black dashed and gray dotted lines indicate $\mathcal{C}_x=1$ and $\mathcal{C}_x=2$, respectively.}
    \label{fig: brokenlaw_concentration}
\end{figure}

\section{Final remarks}
\lac{Discussion}

We have studied the nonlinear evolution of a subdominant ultralight dark matter (ULDM) component with large-amplitude small-scale isocurvature perturbations. Although nonlinear evolution homogenizes the ULDM fraction on galactic scales, ULDM can still dominate the centers of small galaxies by forming an early potential well that seeds subsequent galaxy formation.

The ULDM-dominated region can extend beyond $0.01R_{\rm vir}$ and contribute more than half of the central dark matter density, even when its cosmological fraction is below $\O(10\%)$. Thus, a cosmologically subdominant component can become dynamically important at galaxy centers. In particular, the relative abundance of each dark matter component can depend strongly on its position within a galaxy.

This mechanism may produce either cusp-like or core-like central structures depending on the initial fluctuations. It therefore provides a possible explanation for the observed diversity of galaxy centers within a single multicomponent dark matter scenario.

Our findings motivate several directions for future study:
\begin{itemize}
    \item The first is to determine whether this mechanism can quantitatively account for the observed diversity of galaxy centers.

    \item Rare macroscopic dark matter objects naturally generate stochastic isocurvature perturbations because of their discreteness. Our analysis may therefore also apply to their spatial distribution, for example, to estimate the probability that a supermassive primordial black hole resides near the center of a galaxy.\footnote{We thank Misao Sasaki for useful discussions on primordial black hole scenarios for the origin of supermassive black holes at galactic centers. }

    \item Direct-detection signals in multicomponent dark matter models are usually estimated by assuming that the local fraction of each component follows its cosmological fraction. For example, the signal from axion-like particle (ALP) dark matter is commonly expected to scale as \cite{Sikivie:1983ip}
    \beq
        r_{\rm ALP} g_{\phi\gamma\gamma}^{2},
    \eeq
    where $r_{\rm ALP}$ is the cosmological ALP fraction and $g_{\phi\gamma\gamma}$ is its photon coupling. Our results show that this simple rescaling may fail in the presence of isocurvature perturbations, because the local and cosmological fractions need not coincide. While we have focused on large isocurvature perturbations, it will be important to determine whether mild or small perturbations can also affect the dark matter composition in the Solar neighborhood.
\end{itemize}
\section*{Acknowledgement}
J.N.L.S. acknowledges the support by the European Union and the Czech Ministry of Education, Youth and Sports (Project: MSCA Fellowships CZ FZU III -- \nolinkurl{CZ.02.01.01/00/22_010/0008598}).
W.Y. is supported by JSPS KAKENHI Grant Nos. 22K14029, 23K22486, and 26K00695 and by the Selective Research Fund and Incentive Research Fund from Tokyo Metropolitan University.

\appendix 
\section{Scenarios for small scale isocurvature}
In this appendix we provide some concrete scenarios for the isocurvature perturbation.  

\label{app:1}
\paragraph{Inflationary field fluctuations.}

One possible realization is a coherently displaced scalar field, $\phi$, produced
through the vacuum misalignment mechanism
\cite{Preskill:1982cy,Abbott:1982af,Dine:1982ah}. A sufficiently light
scalar acquires both a homogeneous field value $\bar{\phi}$ within our
observable patch and quantum fluctuations during inflation. The field
begins to oscillate around the minimum of its potential when its mass
becomes comparable to the Hubble parameter and subsequently behaves as
nonrelativistic matter.

For a quadratic potential and a nonzero homogeneous background,
the intrinsic isocurvature perturbation is approximately~\cite{Seckel:1985tj,Lyth:1989pb,Lyth:1991ub}
\beq
S
\simeq
\frac{\delta\rho_u}{\bar{\rho}_u}
\simeq
2\frac{\delta\phi}{\bar{\phi}}.
\eeq
If the scalar mass is negligible and the inflationary Hubble parameter change only
slowly, the field fluctuation is approximately scale invariant. We can 
parametrize the corresponding isocurvature spectrum as
\beq
\mathcal{P}_{SS}^{\rm SI}(k)
=
A_s\beta_{\rm SI}
\left(
\frac{k}{k_*}
\right)^{n_{\rm iso}-1},
\qquad
n_{\rm iso}\simeq 1.
\label{eq:iso_scale_invariant}
\eeq

The spectrum can instead have a strong scale dependence if the effective
mass is efficient and changes during inflation. For example, if the field is
heavy when the CMB modes leave the horizon but becomes light during a later
stage of inflation, the large-scale fluctuations are suppressed relative
to those on smaller scales \cite{Kitano:2023mra}. The homogeneous part $\bar \f$ is also suppressed.

As a simple illustration, consider
\beq
V(\phi,\Phi_{\rm inf})
=
\frac{\lambda}{4}\phi^4
+
\frac{1}{2}M^2(\Phi_{\rm inf})\phi^2,
\qquad
M^2(\Phi_{\rm inf})
=
m_{\phi,0}^2+c\Phi_{\rm inf},
\label{eq:transient_light_potential}
\eeq
where $\Phi_{\rm inf}$ is the slowly rolling inflaton. Suppose that
$M^2(\Phi_{\rm inf})$ changes from negative to positive during inflation.
For $M^2<0$, the scalar is located at the symmetry-breaking minimum,
\beq
\bar{\phi}^{\,2}
=
-\frac{M^2}{\lambda},
\qquad
m_{\rm eff}^2
=
V_{\phi\phi}(\bar{\phi})
=
-2M^2,
\eeq
whereas for $M^2>0$ it is stabilized at the origin and
\beq
\bar{\phi}=0,
\qquad
m_{\rm eff}^2=M^2.
\eeq
The fluctuation mass therefore becomes smaller than the Hubble scale only
near the transition $M^2=0$.

For example, linearizing the inflaton-dependent mass around the transition
as
\beq
\frac{M^2(N)}{H_{\rm inf}^2}
\simeq
\kappa(N-N_0),
\eeq
one obtains
\beq
\frac{m_{\rm eff}^2}{H_{\rm inf}^2}
\simeq
\begin{cases}
2\kappa(N_0-N),
&
N<N_0,
\\[4pt]
\kappa(N-N_0),
&
N>N_0.
\end{cases}
\eeq
Here $N_0$ denotes the time at which $M^2=0$. Since a mode leaving the
horizon at $N=N_k$ satisfies
\beq
\ln\left(\frac{k}{k_0}\right)
\simeq
N_k-N_0,
\qquad
k_0\equiv a(N_0)H_{\rm inf},
\eeq
the condition $m_{\rm eff}^2\lesssim H_{\rm inf}^2$ corresponds
approximately to
\beq
-\frac{1}{2\kappa}
\lesssim
\ln\left(\frac{k}{k_0}\right)
\lesssim
\frac{1}{\kappa}.
\eeq
The resulting field spectrum is therefore a finite, generally asymmetric
bump in $\ln k$, with a characteristic width
\beq
\Delta\ln k
\sim
\mathcal{O}\left(\frac{1}{\kappa}\right).
\eeq
Its precise shape depends on the evolution of $\Phi_{\rm inf}$, the Hubble
parameter, and the post-horizon-exit evolution of the scalar perturbations,
and must in general be obtained by solving the corresponding mode equation.

When the feature is sufficiently narrow compared with the scales resolved
in the subsequent evolution, we phenomenologically approximate the
density-isocurvature spectrum by
\beq
\mathcal{P}_{SS}^{\rm NB}(k)
=
A_s\beta_{\rm NB}
\delta\!\left(
\ln k-\ln k_0
\right).
\eeq
With this convention,\footnote{A narrow field spectrum is mapped directly to a narrow
density-isocurvature spectrum when the density perturbation is linear in
the field fluctuation, as in $S\simeq2\delta\phi/\bar{\phi}$ for
$\bar{\phi}\neq0$. Such a displacement may also be generated after inflation,
for example by a shift of the potential minimum. If instead the abundance is
generated by the fluctuations themselves, $\rho_u\propto\phi^2$, then
$\delta\rho_u(\mathbf{k})$ is a convolution of two field modes,
$
\delta\rho_u(\mathbf{k})
\propto
\int\frac{d^3q}{(2\pi)^3}
\phi(\mathbf{q})\phi(\mathbf{k}-\mathbf{q}),
$
and the density spectrum is broader than the underlying field spectrum.}
\beq
\int d\ln k\,
\mathcal{P}_{SS}^{\rm NB}(k)
=
A_s\beta_{\rm NB}.
\eeq

\paragraph{Incoherent production.}

Another possibility is that the subdominant component is produced
incoherently, for example through the decay of topological defects,
gravitational particle production, or the decay of another particle.
Even when the parent objects or particles do not carry a primordial
isocurvature mode, the finite number of statistically independent
production regions generates an additional stochastic density fluctuation.

If the production mechanism has a finite comoving correlation length of
order $k_0^{-1}$, the dimensional density power spectrum approaches white
noise on larger length scales~\cite{Amin:2022nlh,2503.20881},
\beq
P_S(k)
\longrightarrow
{\rm const.},
\qquad
k\ll k_0.
\eeq
The corresponding dimensionless spectrum therefore behaves as
\beq
\mathcal{P}_{SS}(k)
\propto
k^3,
\qquad
k\ll k_0
.\eeq
This stochastic contribution is an independent isocurvature mode; the
primordial adiabatic perturbation itself does not become isocurvature.

Motivated by this infrared white-noise behavior, we adopt the smooth
broken-power-law template
\beq
\mathcal{P}_{SS}^{\rm BPL}(k)
=
A_s\beta_{\rm BPL}
\frac{1}
{\left(k/k_0\right)^{-3}+(k/k_0)^{p}}.
\eeq
with $p\geq 0$. 
It behaves as
\beq
\mathcal{P}_{SS}^{\rm BPL}(k)
\simeq
A_s\beta_{\rm BPL}
\left(\frac{k}{k_0}\right)^3,
\qquad
k\ll k_0,
\eeq
and approaches the 
\beq
\mathcal{P}_{SS}^{\rm BPL}(k)
\simeq
A_s\beta_{\rm BPL} (\frac{k}{k_0})^{-p},
\qquad
k\gg k_0.
\eeq
The infrared $k^3$ behavior follows from the finite correlation length,
whereas the ultraviolet shape is production-mechanism dependent.
In the numerical simulation we take $p=0$.

\bibliography{bib.bib}
\end{document}